\documentclass{aa}
\usepackage{graphicx}
\usepackage[varg]{txfonts}
\usepackage{natbib}

\newcommand{\zav}[1]{\left(#1\right)}
\newcommand{\hzav}[1]{\left[#1\right]}
\newcommand{\szav}[1]{\left\{#1\right\}}
\newlength\staretab
\newcommand{\Teff}{\ensuremath{T_\mathrm{eff}}}

\newcommand\kms{\ensuremath{\text{km}\,\text{s}^{-1}}}

\begin{document}

\title{Understanding the rotational variability of K2 targets\thanks{Based on
observations collected at the European Southern Observatory, Paranal, Chile (ESO
programme 0103.D-0194(A)).}}
\subtitle{HgMn star KIC 250152017 and blue horizontal branch star KIC 249660366}

\author{J.~Krti\v{c}ka\inst{1} \and A.~Kawka\inst{2} \and 
        Z.~Mikul\'a\v sek\inst{1} \and L.~Fossati\inst{3} \and
        I.~Krti\v ckov\'a\inst{1} \and M.~Prv\'ak\inst{1} \and
        J.~Jan\'\i k\inst{1} \and M.~Skarka\inst{1,4} \and R.~Liptaj\inst{1}}

\institute{Department of Theoretical Physics and Astrophysics,
           Masaryk University, Kotl\'a\v rsk\' a 2, CZ-611\,37 Brno, Czech
           Republic \and
           International Centre for Radio Astronomy Research,
           Curtin University, G.P.O.~Box U1987, Perth, WA 6845, Australia \and
           Space Research Institute, Austrian Academy of Sciences,
           Schmiedlstrasse 6, A-8042 Graz, Austria \and
           Astronomical Institute, Academy of Science of CR, Fri\v{c}ova
           298, CZ-251 65, Ond\v rejov, Czech Republic
}

\date{Received}

\abstract{Ultraprecise space photometry enables us to reveal light variability
even in stars that were previously deemed constant. A large group of such stars
show variations that may be rotationally modulated. This type of light
variability is of special interest because it provides precise estimates of
rotational rates.}{We aim to understand the origin of the light variability of
K2 targets that show signatures of rotational modulation.}{We used
phase-resolved medium-resolution XSHOOTER spectroscopy to understand the light
variability of the stars KIC~250152017 and KIC~249660366, which are possibly
rotationally modulated. We determined the atmospheric parameters at individual
phases and tested the presence of the rotational modulation in the spectra.}{KIC
250152017 is a HgMn star, whose light variability is caused by the inhomogeneous
surface distribution of manganese and iron. It is only the second HgMn star
whose light variability is well understood. KIC 249660366 is a He-weak,
high-velocity horizontal branch star with overabundances of silicon and argon.
The light variability of this star is likely caused by a reflection effect in this
post-common envelope binary.}{}



\maketitle

\section{Introduction}

Rotationally modulated light variability is very common in intermediate-mass
stars \citep[e.g.][]{hummer,mobstersi}. This type of variability is typically
connected with abundance spots in stars without deep subsurface convective
layers \citep[e.g.][]{mythetaaur,prvalis}, but may also have another origin,
for example a magnetically confined circumstellar medium
\citep{labor,towog,magvar}.

In main-sequence stars, abundance spots appear as a result of elemental
diffusion, when the radiatively supported elements emerge on the surface in
enhanced amounts, while the remaining elements sink down due to the effect of
gravity \citep{Mich83,vadog,tuhtalestra}. Abundance spots are particularly
pronounced in stars with strong magnetic fields, in which case Zeeman Doppler
imaging techniques are frequently applied to constrain the spot structure
\citep{piskoc,kowad}.

The derived abundance maps can be used to simulate the light variability and to
determine its origin. The variability of hotter main-sequence, chemically
peculiar stars is typically dominated by the effects of bound-free transitions
of helium and silicon \citep{peter,myhd37776}, while for cooler stars
bound-bound transitions of heavy elements such as iron and chromium play a
dominant role \citep{molnar,seuma}. 

In contrast to stars with a strong magnetic field, according to the classical
picture of non-magnetic, chemically peculiar stars of Am and HgMn types, these
stars do not show any rotational variability, implying an absence of spots. This
picture has been challenged by precise photometric and spectroscopic observations of
these stars \citep{zvezich,adand,morel2014} showing surface spots that may even be evolving \citep{korphiphe}. However, evidence for photometric variability
of these stars was not convincing until space photometry was used
\citep{balpul,paunemag,morel2014,hummerich2018}, which provides brightness
measurements with more than an order of magnitude better precision than
ground-based photometry. \citet{prvphiphe} showed that surface abundance spots
can explain the weak light variability of these stars.

Chemical abundance spots and peculiarities are typically found among
main-sequence stars \citep{niemkepler,ghastat}. However, the effects of
radiative diffusion are not restricted to main-sequence stars, and they
become especially strong in stars with high gravity, that is, hot subdwarfs
and white dwarfs \citep{unbu,unbuhb,vanrimi,miriri}. Subdwarfs typically lack
strong magnetic fields \citep{podland} and the weakness of magnetic field
confinement is possibly the reason for the absence of spots in blue
horizontal branch stars \citep{ernstjakomy}.

\newcommand{\hvezda}{KIC~250152017}
\newcommand{\hvezdad}{KIC~249660366}

\begin{table*}[t]
\centering
\caption{List of spectra used for analysis.}
\label{spek}
\begin{tabular}{lcccccc}
\hline\hline
Star & Arm & Spectrum & JD$-2\,400\,000$ & Phase\tablefootmark{$\ast$} &
Exposure time [s] & S/N\tablefootmark{$\ast\ast$} \\
\hline
\hvezda & UVB & XSHOO.2019-06-24T04:08:45.009 & 58658.67911 & 0.304 & 1100& 90\\
        & VIS & XSHOO.2019-06-24T04:09:32.004 & 58658.67911 & 0.304 & 1006& 50\\
        & UVB & XSHOO.2019-07-31T03:01:21.001 & 58695.63230 & 0.685 & 1100& 80\\
        & VIS & XSHOO.2019-07-31T03:02:08.005 & 58695.63230 & 0.685 & 1006& 50\\
        & UVB & XSHOO.2019-08-02T02:30:22.001 & 58697.61079 & 0.080 & 1100& 90\\
        & VIS & XSHOO.2019-08-02T02:31:09.014 & 58697.61079 & 0.080 & 1006& 50\\
        & UVB & XSHOO.2019-08-10T23:50:26.005 & 58706.49972 & 0.855 & 1100&120\\
        & VIS & XSHOO.2019-08-10T23:51:13.009 & 58706.49972 & 0.855 & 1006& 70\\
\hline
\hvezdad& UVB & XSHOO.2019-07-31T02:27:57.007 & 58695.61142 & 0.694 & 1500& 90\\
        & VIS & XSHOO.2019-07-31T02:28:44.010 & 58695.61142 & 0.694 & 1406& 50\\
        & UVB & XSHOO.2019-08-03T00:17:44.016 & 58698.52100 & 0.778 & 1500& 90\\
        & VIS & XSHOO.2019-08-03T00:18:31.010 & 58698.52100 & 0.778 & 1406& 50\\
\hline
\end{tabular}
\tablefoot{\tablefoottext{$\ast$}{Phase was determined from photometry.}
\tablefoottext{$\ast\ast$}{Signal to noise ratio is given at 4400\,\AA\ and
6500\,\AA\ for the UVB and VIS arms, respectively.}}
\end{table*}

On the other hand, white dwarfs show magnetic fields with diverse strengths
\citep{valy,asb,landbt}, but indications of surface abundance spots in white
dwarfs are scarce \citep{dupchaven,uhlprom}. Therefore, we took the list of
white dwarfs observed by the Kepler satellite \citep{hermes}, extracted those
that most likely have significant rotational variability, and performed a
detailed spectroscopic study to identify their nature. Here, we report the first
results of this survey that led us to conclude that two previously considered
white dwarfs possibly presenting rotational modulation are in reality less
evolved stars: a HgMn star and a He-weak blue horizontal branch star.

\section{Observations and spectral analysis}

We obtained spectra of \hvezda\ and \hvezdad\ as part of the ESO proposal
0103.D-0194(A). The spectra were acquired with the XSHOOTER spectrograph
\citep{xshooter} mounted on the 8.2m UT2 Kueyen telescope and these observations
are summarized in Table~\ref{spek}. The spectra were obtained with the UVB and
VIS arms providing an average spectral resolution ($R=\lambda/\Delta\lambda$) of
9700 and 18\,400, respectively. Although medium-resolution spectra are not ideal
for abundance analysis, the abundance determination is based on multiple strong
lines for most elements. This mitigates the disadvantages of the
medium-resolution spectra and enables us to use even medium-resolution spectra
for abundance analysis \citep[e.g.][]{kawnltt,gvarpol}. As a result, the
difference due to the spectral resolution mostly affects the number of elements
one can measure and with it the precision of those measured elements whose
lines are blended by the elements that have not been measured. The calibrated
spectra were extracted from the European Southern Observatory (ESO) archive. We determined the radial velocity
from each spectrum by means of a cross-correlation function using the
theoretical spectrum as a template \citep{zvezimi}, and shifted the spectra to
the rest frame.

We used simplex minimization to determine stellar parameters \citep{kobr}. The
analysis was performed in two steps. In the first step, we determined the effective
temperature $T_\text{eff}$ and surface gravity $\log g$ by fitting the observed
spectra with fixed abundances. In the second step, we fixed the effective
temperature and surface gravity and determined individual abundances relative to
hydrogen $\varepsilon_\text{el}=\log(n_\text{el}/n_\text{H})$ again by fitting
the observed spectra. We analysed the normalized spectra, but to test the
derived effective temperatures and surface gravities we additionally also fitted
flux calibrated spectra. To this end and to account for possible systematics in
the absolute flux calibration, we introduced a parameter that rigidly scaled the
synthetic fluxes in the fitting procedure to the observed fluxes.

We processed each spectrum separately. For the analysis of \hvezda, we used
ATLAS12 \citep{atlas2,atlas3} local thermodynamical equilibrium (LTE) model
atmospheres. We applied the BSTAR2006 grid \citep{bstar2006} and NLTE (non-LTE)
TLUSTY models
to determine the parameters of \hvezdad. The spectrum synthesis was done using the
SYNSPEC code \citep{bstar2006}, neglecting isotopic splitting.

\begin{table}[t]
\caption{Basic astrometric and photometric parameters of studied stars.}
\label{hveztab}
\centering
\begin{tabular}{lcc}
\hline
KIC (EPIC) & 250152017 & 249660366 \\
\hline
$\alpha$ (J2000)   & 15h 25m 08.282s & 15h 24m 48.378s\\
$\delta$ (J2000)   & $-13^\circ$ $34'$ $49.11''$ & $-19^\circ$ $44'$ $53.62''$\\
$\pi$ [mas]        & $0.296\pm0.057$       & $0.439\pm0.074$ \\
$d$ [kpc]          & $3.2\pm0.7$           & $2.2\pm0.4$ \\
$m_V$ [mag]        & $14.726\pm0.042$      & $14.967\pm0.045$\\
$m_B$ [mag]        & $14.769\pm0.040$      & $15.014\pm0.030$\\
$E(B-V)$ [mag]     & $0.12\pm0.02$         & $0.09\pm0.02$\\
$U,\,V,\,W$ [\kms] & $[-10,\,-174,\,-23]$  & $[22,\,-205,\,92]$ \\
\hline
\end{tabular}
\end{table}

Table~\ref{hveztab} summarizes the available astrometric and photometric data for
both stars. The coordinates and magnitudes were derived from the Mikulski
Archive for Space Telescopes (MAST) K2
catalogue \citep{epic}, while the parallax $\pi$ and proper motions are taken
from the GAIA data release 2 (DR2) data \citep{gaia1,gaia2}. The distances $d$ were determined
from the parallax using the method described by \cite{luri}. The colour excesses were
determined from Galactic reddening maps \citep{prasan} for locations
corresponding to the stars. The space velocities in Table~\ref{hveztab} were
determined from GAIA DR2 proper motions and radial velocities using
\citet{vlapoh}.

\section{Chemically peculiar star \hvezda}

\begin{table}[t]
\caption{Derived mean parameters of the studied stars compared to solar
abundances \citep{asp09}.}
\label{hvezpar}
\centering
\begin{tabular}{lccc}
\hline
KIC & 250152017 & 249660366 & Sun \\
\hline
$T_\text{eff}$ [K]      & $12160\pm130$ & $19700\pm300$\\
$\log g$ [cgs]          & $3.95\pm0.02$ & $4.57\pm0.06$\\
$M$ $[M_\odot]$         & $3.33\pm0.08$ & $0.27\pm0.10$\\
$R$ $[R_\odot]$         & $3.19\,\pm0.10$ & $0.45\pm0.09$\\
$v\sin i$ [\kms]        & $22\pm5$    & $26\pm4$\\
$v_\text{rad}$ [\kms]   & $2\pm4$       & $107\pm3$ \\
$\log\varepsilon_\text{He}$ & $-1.60 \pm 0.06$ & $-2.91 \pm 0.26$ & $-1.07$\\
$\log\varepsilon_\text{C}$  & $-4.08 \pm 0.04$ & $-4.47 \pm 0.18$ & $-3.57$ \\
$\log\varepsilon_\text{N}$  &                  & $-4.37 \pm 0.01$ & $-4.17$\\
$\log\varepsilon_\text{O}$  &                  & $-4.40 \pm 0.49$ & $-3.31$\\
$\log\varepsilon_\text{Mg}$ & $-5.04 \pm 0.03$ & $-5.06 \pm 0.08$ & $-4.40$ \\
$\log\varepsilon_\text{Al}$ &                  & $-6.39 \pm 0.05$ & $-5.55$ \\
$\log\varepsilon_\text{Si}$ & $-5.02 \pm 0.03$ & $-4.05 \pm 0.07$ & $-4.49$ \\
$\log\varepsilon_\text{S}$  &                  & $-5.50 \pm 0.05$ & $-4.88$\\
$\log\varepsilon_\text{Ar}$ &                  & $-4.86 \pm 0.02$ & $-5.60$ \\
$\log\varepsilon_\text{Ca}$ & $-6.02 \pm 0.22$ & $-6.32 \pm 0.02$ & $-5.66$ \\
$\log\varepsilon_\text{Ti}$ & $-6.44 \pm 0.05$ &  & $-7.05$ \\
$\log\varepsilon_\text{Mn}$ & $-4.61 \pm 0.04$ &  & $-6.57$ \\
$\log\varepsilon_\text{Fe}$ & $-4.69 \pm 0.02$ &  & $-4.50$ \\
$\log\varepsilon_\text{Ni}$ & $-6.92 \pm 0.27$ &  & $-5.78$ \\
$\log\varepsilon_\text{Y}$  & $-7.87 \pm 0.07$ &  & $-9.79$ \\
$\log\varepsilon_\text{Hg}$ & $-5.08 \pm 0.04$ &  & $-10.83$ \\
\hline
\end{tabular}
\end{table}

\begin{figure*}[t]
\centering
\includegraphics[width=\hsize]{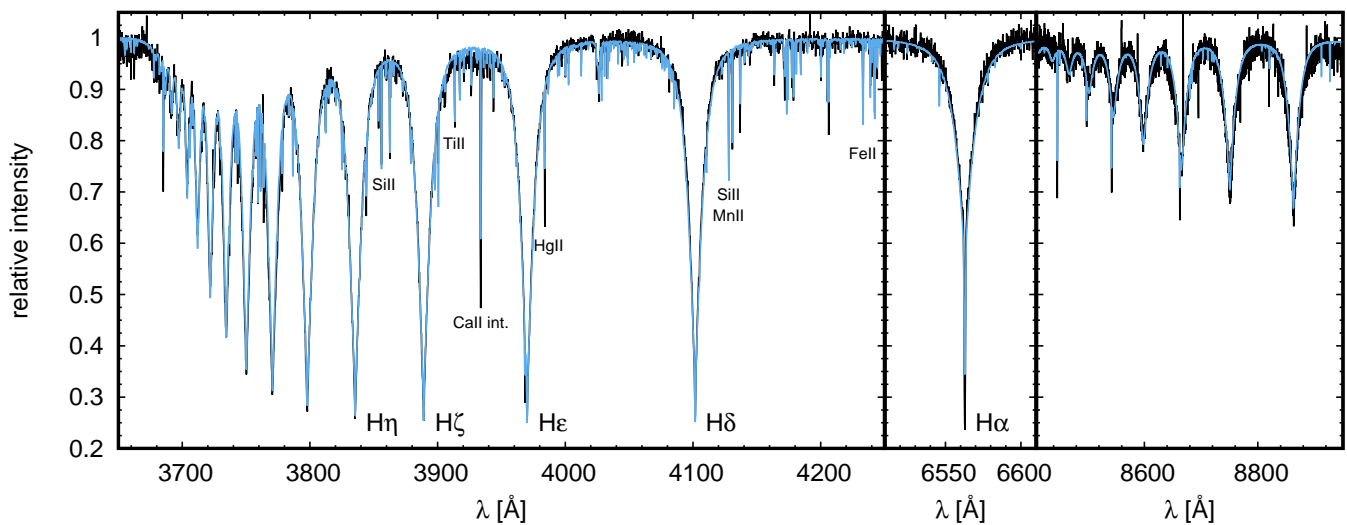}
\caption{Comparison of observed (black line) and best fit synthetic (blue line)
spectra of \hvezda.}
\label{hvezdaspek}
\end{figure*}

The phases of \hvezda\ spectra in Table~\ref{spek} were determined with
ephemeris (in barycenter corrected Julian date)
\begin{equation}
\label{hvezdaef}
\text{JD}=2\,458\,031.346(21)+5.0065(15)E
\end{equation}
derived from K2 photometry\footnote{K2 photometry was obtained from the MAST
archive http://archive.stsci.edu, \citet{epic}.}
\citep[see][for details]{mikmon}. Mean phase averaged parameters and their
uncertainties derived from fitting the individual observed spectra by LTE models
(see Fig.~\ref{hvezdaspek}) are given in Table~\ref{hvezpar}. The strongest
lines used for the abundance determination are listed in Table~\ref{hvezdacar}.

\begin{table}[t]
\caption{List of wavelengths (in \AA) of the strongest lines used for \hvezda\
abundance determination.}
\label{hvezdacar}
\begin{tabular}{ll}
\hline
\ion{He}{i}  & 3819.603, 3819.614, 4026.187, 4026.198, 4026.199, \\
             & 4026.358, 4143.761, 4387.929, 4471.473, 4471.485, \\
             & 4471.488, 4471.682, 4713.139, 4713.156, 4921.931, \\
             & 5015.678 \\
\ion{C}{ii}  & 3920.681, 4267.001, 4267.183 \\
\ion{Mg}{ii} & 4384.637, 4390.572, 4433.988, 4481.126, 4481.150, \\
             & 4481.325 \\
\ion{Si}{ii} & 3853.665, 3856.018, 3862.595, 4128.054, 4130.872, \\
             & 4130.894, 5041.024, 5055.984, 5056.317 \\
\ion{Ca}{ii} & 3706.024, 3736.902 \\
\ion{Ti}{ii} & 3685.189, 3685.204, 3706.216, 3741.635, 3757.688, \\
             & 3759.296, 3761.323, 3761.883, 3913.468, 4163.648, \\
             & 4290.219, 4294.099, 4300.049, 4301.914, 4307.863, \\
             & 4312.864, 4314.975, 4395.033, 4399.772, 4443.794, \\
             & 4468.507, 4501.273, 4533.969, 4549.617, 4563.761, \\
             & 4571.968, 5226.543 \\
\ion{Mn}{i}  & 4033.062, 4041.355 \\
\ion{Mn}{ii} & 3715.269, 3743.382, 3812.239, 3812.524, 3843.886, \\
             & 3844.161, 3878.992, 3897.604, 3898.056, 3917.318, \\
             & 3943.598, 3943.858, 4128.129, 4136.902, 4174.318, \\
             & 4184.454, 4200.270, 4205.381, 4206.367, 4238.791, \\
             & 4239.187, 4242.333, 4251.717, 4252.963, 4253.025, \\
             & 4253.112, 4259.200, 4282.490, 4308.158, 4326.639, \\
             & 4356.621, 4363.255, 4365.217, 4379.669, 4434.067, \\
             & 4441.991, 4478.637, 4500.543, 4518.956, 4749.112, \\
             & 4764.728, 4770.351, 4784.625, 4791.782, 4806.823, \\
             & 5123.327, 5295.384, 5295.412, 5297.000, 5297.028, \\
             & 5297.056, 5299.302, 5299.330, 5299.386, 5302.402, \\
             & 5302.431 \\
\ion{Fe}{ii} & 4173.461, 4178.862, 4233.172, 4303.176, 4385.387, \\
             & 4416.830, 4491.405, 4508.288, 4515.339, 4520.224, \\
             & 4522.634, 4549.474, 4555.893, 4583.837, 4629.339, \\
             & 4923.927, 5001.959, 5018.440, 5100.607, 5100.727, \\
             & 5100.852, 5169.033, 5197.577, 5227.481, 5234.625, \\
             & 5260.259, 5276.002, 5316.615 \\
\ion{Ni}{ii} & 4067.031 \\
\ion{Y}{ii}  & 4374.935, 4883.684 \\
Hg           & 3983.890, 4046.559 \\
\hline
\end{tabular}
\end{table}

\begin{figure}[t]
\centering
\includegraphics[width=\hsize]{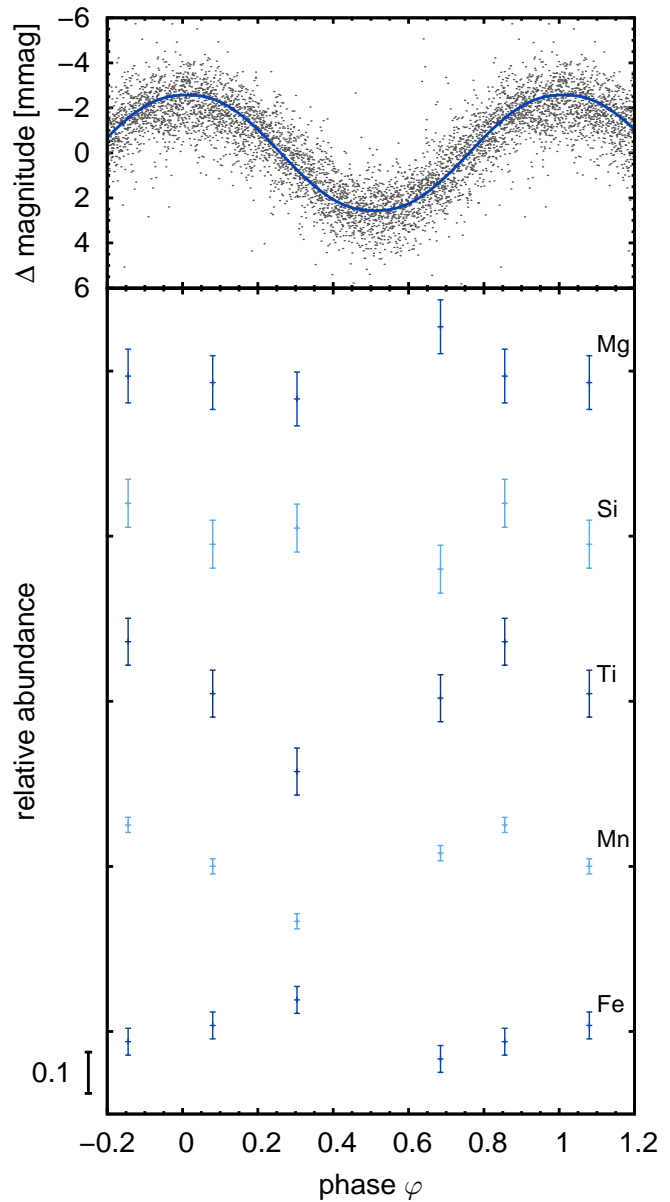}
\caption{Phase variations of \hvezda. {\em Upper panel}: K2 magnitudes. The solid
line denotes a light curve simulated assuming surface abundance spots of manganese
and iron corresponding to the observed spectral variations. {\em Bottom panel}:
Relative variations of abundances. Uncertainties were determined following
\citet{chlapec}.}
\label{250152017prom}
\end{figure}

\begin{figure}[t]
\includegraphics[width=\hsize]{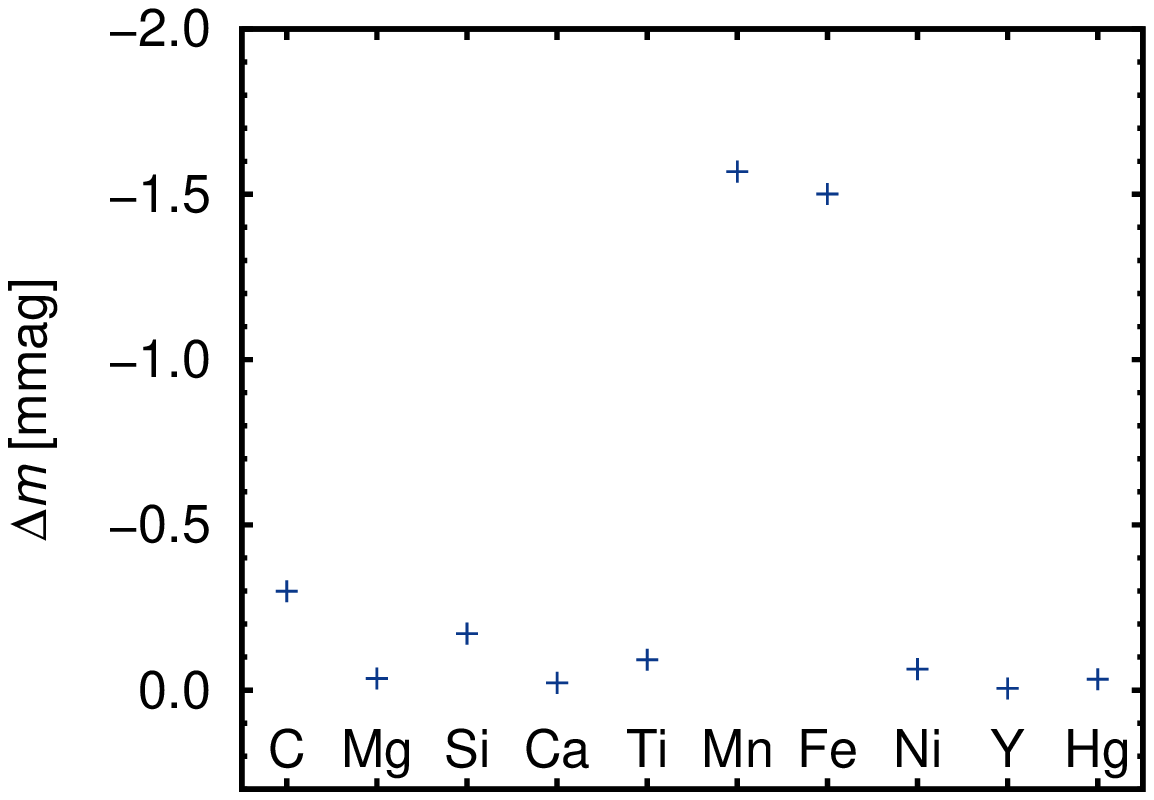}
\caption{Relative magnitude differences between flux at $6000\,\AA$ calculated
with enhanced abundance of a given element and flux at \hvezda\ abundances.}
\label{ktereprv11g}
\end{figure}

Judging from the effective temperature $T_\text{eff}=12160\pm130\,\text{K}$ and surface
gravity $\log g=3.95\pm0.02$,  \hvezda\ is a main-sequence star.
An underabundance of He, C, Mg, and Si and an overabundance of Mn, Y, and Hg indicates
that it belongs to the HgMn group of chemically peculiar stars
\citep[CP3,][]{preston,maitzen,ghakat}. Mercury typically shows isotopic
splitting in HgMn stars \citep{bilartut,woohg,ares} possibly affecting the
derived abundance. We used the BONNSAI\footnote{The Bonn Stellar Astrophysics Interface web service is available
at https://www.astro.uni-bonn.de/stars/bonnsai.} service \citep{bonnsai} to
determine the evolutionary parameters of the star from observational parameters
assuming solar metallicity evolutionary models \citep{brott}. We derived a mass of
$M=3.33\pm0.08\,M_\odot$, a radius of $R=3.19\,\pm0.10\,R_\odot$, and an age of
$162\pm10\,$Myr. With a known projection of rotational velocity of $v\sin
i=22\pm5\,\kms$ and assuming that the period given in Eq.~\eqref{hvezdaef} is
due to rotation, we determined the inclination as $i=42^\circ\pm15^\circ$.

In addition, we determined the stellar parameters from flux calibrated data. The
derived parameters $T_\text{eff}=12\,500\pm100\,\text{K}$ and  $\log
g=3.92\pm0.01$ reasonably agree with values determined from normalized spectra.

The parameters derived from evolutionary tracks disagree with GAIA DR2 data.
From a distance of $3200\pm700\,$pc, the visual magnitude is
$m_V=14.726\pm0.042\,$mag, the colour excess is $E(B-V)=0.12\pm0.02$, the bolometric
correction is $BC=-0.72\pm0.03\,$mag \citep[see also
\citealt{spravnybyk}]{bckytka}, the deredened magnitude is
$V=14.35\pm0.07\,\text{mag,}$ and the absolute bolometric magnitude is
$1.1\pm0.5\,$mag. This is significantly higher than the value derived from the
evolutionary tracks, $-1.0\pm0.1\,$mag. Using the effective temperature from
spectroscopy, this implies a lower stellar radius of $R=1.2\pm0.3\,R_\odot$ than
that derived from the evolutionary tracks. A lower radius gives a rotational
velocity of $v_\text{rot}=12\pm3\,\kms$, which is significantly lower than the
spectroscopically derived value. This most likely implies that the GAIA DR2 distance
is incorrect, possibly due to the binary nature of the object.

The presence of a companion is common in HgMn stars
\citep[e.g.][]{ryabdvoj,niemhgmn}, but we have not detected radial velocity
variations larger than about $10\,\kms$ in \hvezda. Neither have we detected any
features in the spectra that can be attributed to the companion. Therefore the
maximum flux ratio of both components is about 0.1 in the VIS arm. This implies
that the companion could be a main-sequence star of spectral type F5 or later
\citep[using main-sequence parameters from][]{har}. Such a star would cause
radial velocity variations below the detection limit even assuming tidal
locking. Large $V$ velocity (see Table~\ref{hveztab}) would imply that the
object is either a member of the halo population or that it is a runaway star, however
this is questionable given the problem with parallax.

Figure~\ref{250152017prom} shows that abundances of individual elements
have phase-dependent variations. The phase variations of individual elements are
mutually shifted. However, there seems to be a general trend that the heavier
elements Si, Ti, Mn, and Fe vary in phase with the light curve, while Mg
possibly varies in antiphase.

The periodic line profile variations in chemically peculiar stars are
interpreted as a result of surface abundance spots
\citep[e.g.][]{leuma,marnyrus}, which also cause photometric variations
\citep{myhd37776,seuma}. Precise photometry of HgMn stars also reveals
rotational variability \citep{morel2014,strassmeier2017,hummerich2018}. However, evidence of abundance spots on the surface of HgMn stars is scarce
\citep{kochukhov2007,briquet2010,hubrig2010,makaganiuk2011}. The photometric
variability in HgMn stars can be also attributed to flux redistribution due to
modified opacity in abundance spots \citep{prvphiphe}.

To identify which elements may contribute to the light variability, in
Fig.~\ref{ktereprv11g} we plot the relative magnitude difference $\Delta m =
-2.5\log(F_\text{el}/F_\ast)$ between the flux $F_\text{el}$ calculated with
the abundance of a given element multiplied by a factor of 1.1, and the flux $F_\ast$
calculated for the chemical composition of  \hvezda\ . The adopted multiplicative factor 1.1
corresponds to typical abundance variations found in Fig.~\ref{250152017prom}.
The fluxes used in Fig.~\ref{ktereprv11g} were smoothed by a Gaussian filter
with a dispersion of $100\,\AA$ and evaluated at wavelength 6000\,\AA, which
corresponds to the maximum of the Kepler response function. From
Fig.~\ref{ktereprv11g} it follows that mostly manganese and iron contribute to the
light variability of \hvezda.

We simulated light variations of \hvezda\ assuming that one hemisphere has
a greater abundance of Mn and Fe, higher than the other hemisphere by
$\Delta\varepsilon_\text{Mn}=0.08$ and $\Delta\varepsilon_\text{Fe}=0.05$,
respectively. These values correspond to the amplitudes of the abundance
variations determined from the spectra in individual phases
(Fig.~\ref{250152017prom}, bottom panel). We calculated the synthetic spectra
for individual abundances and integrated the specific intensities over the
visible stellar surface and over the instrumental response curve \citep[see][for
more details of the code]{mythetaaur}. We adopted a fit to the instrument
response curve as derived from the Kepler Instrument Handbook \citep{kepinst} and
given in the Appendix~\ref{slepak}. The predicted and observed light variations
in Fig.~\ref{250152017prom} (upper panel) reasonably agree demonstrating that
surface abundance variations of Mn and Fe are able to explain the observed light
variability of \hvezda.


\section{He-poor high-velocity star \hvezdad}

\begin{figure}[t]
\centering
\includegraphics[width=\hsize]{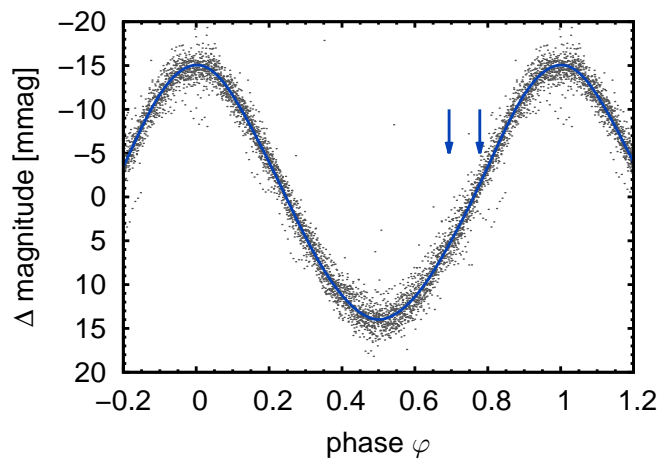}
\caption{Light variations of \hvezdad . The solid blue line denotes a fit by the spot model
of \citet{prvakphd}. The blue arrows correspond to the phases when the spectra in
Table~\ref{spek} were taken.}
\label{249660366prom}
\end{figure}

\begin{figure*}[t]
\centering
\includegraphics[width=\hsize]{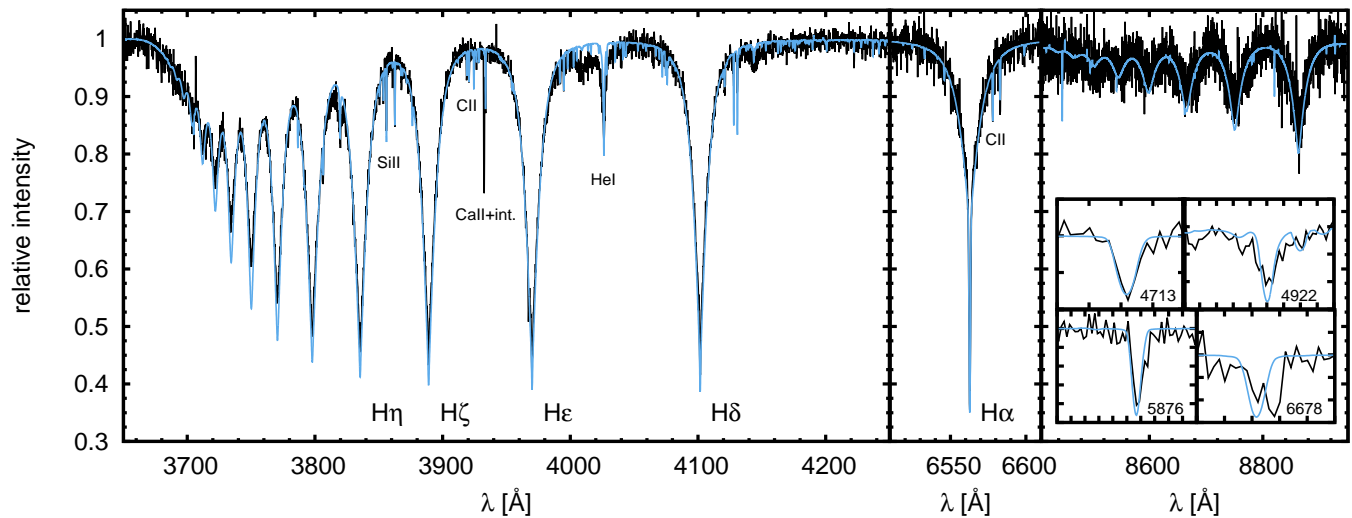}
\caption{Comparison of observed (black line) and best fit synthetic (blue line)
spectra of \hvezdad. The inset in the right plot shows selected \ion{He}{i}
lines. The vertical and horizontal scales of the inset are 0.05, and 1\,\AA,
respectively.}
\label{249660366sp}
\end{figure*}

\begin{figure}[t]
\centering
\includegraphics[width=\hsize]{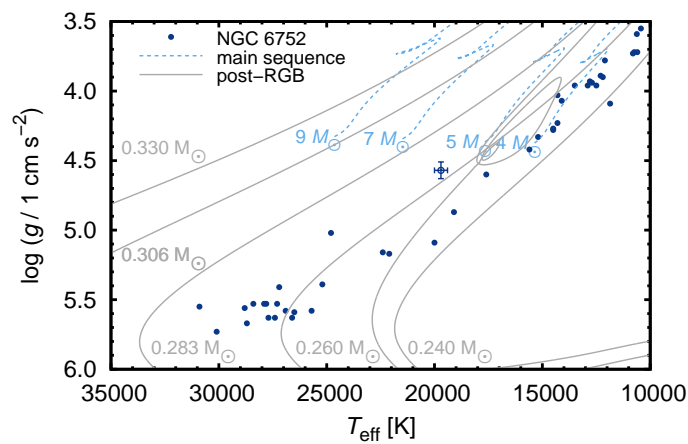}
\caption{Location of \hvezdad\ in $\log g$ -- $\Teff$ diagram (blue cross).
Overplotted are the main-sequence evolutionary tracks \citep{sylsit}, post-red giant branch evolutionary tracks \citep{hall8000}, and the parameters of the blue
horizontal branch stars from NGC~6752 \citep{moni,moelopra}.}
\label{249660366gt}
\end{figure}

The phases of \hvezdad\ spectra in Table~\ref{spek} were determined using the
ephemeris (in barycenter corrected Julian date)
\begin{equation}
\label{hvezdadper}
\text{JD}=2\,458\,033.2256(21)+2.68505(12)E,
\end{equation}
which was derived from K2 photometry (see Fig.~\ref{249660366prom}). Mean (phase averaged) parameters of \hvezdad\ (Table~\ref{hvezpar}) derived from
fitting the individual spectra by NLTE models (Fig.~\ref{249660366sp}) place the
object below the solar metallicity main sequence (Fig.~\ref{249660366gt}). The
parameters derived from fitting flux calibrated spectra give a slightly higher
effective temperature $T_\text{eff}=20\,600\pm200\,$K and similar surface
gravity $\log g=4.64\pm0.08$ compared to the parameters derived from normalized
spectra (Table~\ref{hvezpar}). The star shows an underabundance of helium and
light elements, and an overabundance of silicon and argon (see
Table~\ref{hvezdadcar} for the list of analysed lines). With
$m_V=14.967\pm0.045\,\text{mag}$ and $E(B-V)=0.09\pm0.02$, the deredened
$V=14.69\pm0.08\,\text{mag}$. With the GAIA DR2 distance of $2200\pm400\,$pc, the
absolute magnitude is $M_V=2.98\pm0.40\,$mag, which with the bolometric
correction $BC=-1.86\pm0.03\,$mag from \citet[see also
\citealt{spravnybyk}]{bckytka} and with spectroscopically determined parameters
gives a stellar radius and mass of $R=0.45\pm0.09\,R_\odot$ and
$M=0.27\pm0.10\,M_\odot$, respectively. These parameters are typical for
subdwarfs and blue horizontal branch stars \citep{heberpreh}.


The \hvezdad\  light curve (Fig.~\ref{249660366prom}) could be explained either
by binary effects, by pulsations, or by the abundance spots. If the light
variations are due to ellipsoidal variability (distortion of the stellar
surface), then the time of the observation should correspond to a period of
relatively large changes in the radial velocity. The orbital period would be
twice that determined in Eq.~\eqref{hvezdadper}. Because there are no lines of
the secondary star present in the spectra, we can assume that the total mass of
the system is that of a putative primary. By using Kepler's third law, we
calculated the semi-major axis as $8.6\,R_\odot$. We used our binary code that
simulates the light curve assuming stellar distortion within the Roche model
(adopting a companion mass of $0.2\,M_\odot$) to predict the amplitude of the
light variability. The derived amplitude is two orders of magnitude lower than
the observed amplitude of the light variability. Subdwarfs are frequently found
in binaries with degenerate components \citep{heberpreh}. However, adopting
a higher companion mass would imply larger binary separation and even lower
amplitude of the light variability. Thus, we conclude that ellipsoidal
variations due to a subdwarf component are unlikely to cause light variability in
\hvezdad.

\begin{table}[t]
\caption{List of wavelengths (in \AA) of the strongest lines used for \hvezdad\
abundance determination.}
\label{hvezdadcar}
\begin{tabular}{ll}
\hline
\ion{He}{i}  & 3867.470, 3867.482, 3926.535, 4713.139, 4713.156, \\
             & 4921.931, 5015.678, 5875.614, 5875.615, 5875.625, \\
             & 5875.640, 5875.966, 7065.176, 7065.214, 7065.707 \\
\ion{C}{ii}  & 3918.968, 3920.681, 4267.001, 6578.050, 6582.880\\
\ion{N}{ii}  & 3994.997, 4607.153, 4621.393, 4630.539, 4643.086, \\
             & 5001.134, 5001.474, 5005.150 \\
\ion{O}{ii}  & 4414.899, 4649.135 \\
\ion{Mg}{ii} & 4481.126, 4481.150, 4481.325 \\
\ion{Al}{iii}& 4479.885, 4529.189 \\
\ion{Si}{ii} & 3853.665, 3856.018, 3862.595, 4621.418, 4621.722, \\
             & 6347.109 \\
\ion{Si}{iii}& 4552.622, 4567.840, 4574.757 \\
\ion{S}{ii}  & 4924.110, 4925.343, 5009.567 \\
\ion{Ar}{ii} & 3850.581, 4379.667, 4426.001, 4430.189, 4430.996, \\
             & 4579.349, 4609.567, 4657.901, 4726.851, 4735.906, \\
             & 4764.864, 4806.021, 4965.060 \\
\ion{Ca}{ii} & 3933.663 \\
\hline
\end{tabular}
\end{table}

The light variations in overcontact binaries may also produce light curves that
have a nearly sinusoidal shape \citep{tylmer}. In this case the true period would
again be twice that determined in Eq.~\eqref{hvezdadper} and assuming an equal mass
of each star of $0.3\,M_\odot$ gives an orbital separation of $11\,R_\odot$,
which is significantly larger than the radius of a subdwarf. Consequently,
the model of an overcontact binary also does not provide a reliable explanation
of the \hvezdad\ light curve.

A large fraction of subdwarfs show light variations due to a reflection effect on
a cool companion \citep{scherebos}. Detailed modelling of such an effect is
complex \citep[e.g.][]{budref}, so we just estimated the amplitude of expected
light variations. Using $R_1$, $R_2$, $T_1$, and $T_2$ to denote the radii and
effective temperatures of both components, the radiative power acquired by the
companion from the Stefan-Boltzmann law $\sigma T_1^4 \pi R_2^2 R_1^2/a^2$ leads
to an increase in the companion effective temperature of
$T_2\hzav{1+\zav{T_1/T_2}^4 \zav{R_1/(2a)^2}}^{1/4}$. Using the Rayleigh–Jeans
law, the amplitude of the magnitude change due to the reflection effect is then
\begin{equation}
\label{dm}
\Delta m\approx\frac{R_2^2T_2}{R_1^2T_1}
\szav{\hzav{1+\zav{T_1/T_2}^4 \zav{R_1/(2a)^2}}^{1/4}-1}.
\end{equation}
With orbital separation $a$ derived from Kepler's third law as a function of
secondary mass, Eq.~\ref{dm} gives the amplitude of the reflection effect as a
function of the parameters of the secondary. We calculated the magnitude of the
reflection effect using the main-sequence stellar parameters of \citet{har} for a
wide range of companions with $T_2$ down to $3\,$kK. This analysis shows that a
low-mass companion with $M_2\lesssim0.5\,M_\odot$ is able to cause the observed
light variability due to the reflection effect and still its radiation will not
affect the combined spectrum of the binary.  The object shows only a marginal
change in the radial velocities $\Delta v_\text{rad}=-3\pm6\,\kms$ between the
two spectra. This does not contradict the expected radial velocity curve in the
case of a reflection effect, which should show maximum around phase 0.75, as
follows from Fig.~\ref{249660366prom}. We checked the presence of a putative
companion using the spectral energy distribution constructed using the VOSA tool
\citep{vosa}. The observed flux limits the effective temperature of the
main-sequence companion $T_2\leq4\,$kK, which agrees with previous constraints.
Consequently, the reflection effect provides plausible explanation of the 
\hvezdad\ light curve.

The light variations could also be caused by pulsations. The location of
\hvezdad\ in the Hertzsprung-Russell diagram is close to the region populated by
$\beta$~Cep variables and by slowly pulsating B stars \citep{valcik}. However,
subdwarfs typically pulsate with periods of the order of a fraction of a day
\citep{ostrofont,kilkecpul,ostensil,kapitool}, which is much shorter than the
period found in \hvezdad. Moreover, such stars typically show several
pulsational periods, which is not the case for \hvezdad. The studied star is
slightly cooler than blue large-amplitude pulsators \citep{blap}. These are
expected to be post-common-envelope objects on their way towards the white dwarf
evolutionary stage, with pulsations driven by the opacity of iron-group elements
\citep{basnici}. Also these objects pulsate with significantly shorter periods
(of the order of tens of minutes) than found in \hvezdad. Consequently, it is
not likely that the light variations of \hvezdad\ originate from pulsations. 

The remaining possibility is that the light variability is caused by surface
spots, which are connected with peculiar abundances in hot stars
\citep[e.g.][]{leuma,marnyrus}. To search for a possible source of light
variations, we calculated additional atmosphere models with an abundance of silicon
and argon three times higher and ten times higher than the solar one,
respectively. The enhancement of the abundance leads to an increase in the optical flux
at $6000\,$\AA\  by about 0.01\,mag in the case of silicon and by 0.05\,mag in the
case of argon, which would explain the observed light variations.

We used the code of \citet{prvakphd} to test if the observed light variations
can be reproduced by surface spots. The code uses procedure inspired by genetic
algorithms to search for a surface brightness distribution that fits the
observed light curve best. We assumed a maximum intensity contrast of 7\%. The
code fits the observed photometric variability using a large bright surface spot
located at latitudes that nearly directly face the observer. The fit in
Fig.~\ref{249660366prom} nicely reproduces the observed light curve.

If the light variations are caused by surface spots, then the ephemeris of
Eq.~\eqref{hvezdadper} would give the period of rotation, which in combination
with the radius from Table~\ref{hvezpar} gives the rotational velocity
$9\pm2\,\kms$. This value is lower than the measured value of $v\sin i$, which
is $26\pm4\,\kms$. Consequently, the rotational modulation as a source of light
variability is unlikely to leave only ellipsoidal variations as the remaining
plausible explanation of the observed light variations.

Chemical peculiarities are frequently observed in horizontal branch stars
\citep{edchem,nemgal,gechem,esosubwind} and are interpreted as being a result of
diffusion processes caused by competing radiative and gravity accelerations
\citep{unbu,vanrimi,miriri,hu}. Similar peculiarities are also found in helium-weak stars \citep{ghastat}, which are typically magnetic \citep{shumagrot}. The
absence of a strong surface magnetic field possibly explains why the search
for rotational variability in horizontal branch stars was negative
\citep{ernstjakomy}.

Some silicon lines in the spectra require a lower abundance than given in
Table~\ref{hvezpar} to derive a precise fit. Moreover, even with NLTE models we
were not able to reproduce the strong helium lines with any abundance. These
features most likely point to vertical abundance gradients in the atmosphere,
which are frequently found in chemically peculiar stars
\citep{lela,ryabca,vesel}. Some helium lines (4388\,\AA\ and 6678\,\AA) are
split due to isotopic splitting (see the inset of Fig.~\ref{249660366sp}), which
is also a common feature in peculiar stars \citep{saju,harthetri}. We have
excluded these lines in the abundance analysis.

The object appears at a relatively high Galactic latitude of $30.2^\circ$.
Therefore, the star is located about 1.1\,kpc above the Galactic plane. The
location in the Galaxy and high spatial velocities (Table~\ref{hveztab}) suggest
that \hvezdad\ likely belongs to Pop II. The position of the star in the
$T_\text{eff}$ versus $\log g$ diagram (Fig.~\ref{249660366gt}) and its mass
correspond to post-red giant branch stars after the common-envelope phase. The presence of a close
companion deduced from photometric variations is consistent with this
evolutionary scenario.
Consequently, we conclude that \hvezdad\ is a high-velocity horizontal branch star
whose chemical composition is the result of gravitational settling, radiative
diffusion, and possibly of galactic abundance gradients. The star was likely
stripped off its envelope by its companion \citep{hall8000,basnici} and
is currently evolving towards the white dwarf stage.

\section{Conclusions}

We studied the photometric variability of the two K2 targets \hvezda\ and \hvezdad\
to test the rotational modulation and the nature of the light curves. We derived
phase-resolved spectroscopy of both stars and determined parameters from
spectroscopic analysis as a function of rotational phase. The selected stars
were included among white dwarfs observed by the Kepler satellite
\citep{hermes}, but we have shown that neither star is a white dwarf.

Star \hvezda\ shows an overabundance of the heavy elements titanium, manganese,
yttrium, and mercury, and an underabundance of helium, carbon, magnesium, and
silicon. These are typical features of HgMn (CP3) stars that appear due to
processes of selective radiative force and gravitational settling. The phase-resolved spectroscopy of \hvezda\ reveals variations of titanium, manganese, and
iron lines, which are consistent with their rotational origin. Surface abundance
spots of manganese and iron are able to explain the rotational variability of
the star. This is the second well-documented study on the nature of light
variability of HgMn stars after \citet{prvphiphe}.

\hvezdad\ is helium-poor star that shows an overabundance of silicon and argon, low
mass, and high gravity corresponding to horizontal branch stars. The peculiar
abundance could be also attributed to diffusion processes. Although the
phase coverage of the spectra does not allow for a detailed study of spectral
variability, the analysis shows that the reflection effect due to the unseen
companion provides the most likely explanation of the light variability of this
system, which we expect to be of post-common envelope type.

\begin{acknowledgements}
The authors thank Dr. Ernst Paunzen for his comments. This research was
supported by grant GA\,\v{C}R 18-05665S. MS acknowledges the financial support
of the Operational Program Research, Development and Education -- Project
Postdoc@MUNI (No. CZ.02.2.69/0.0/0.0/16\_027/0008360). This paper includes data
collected by the K2 mission. Funding for the K2 mission is provided by the NASA
Science Mission directorate. Computational resources were supplied by the
project "e-Infrastruktura CZ" (e-INFRA LM2018140) provided within the program
Projects of Large Research, Development and Innovations Infrastructures. 
\end{acknowledgements}

\bibliographystyle{aa}
\bibliography{papers}

\appendix
\section{Fit of the Kepler instrumental profile}
\label{slepak}

We fit to the instrument response curve derived from the Kepler Instrument
Handbook \citep{kepinst} by a curve
\begin{equation}
\Phi(\lambda)=\left\{\begin{array}{cl}
      \exp\szav{-x^2\hzav{a_0+x^4c_0}}, & x<0,\\
      \exp\szav{-x^2\hzav{a_1+x^2(b_1+x^2c_1)}}, & x\geq0,\\
              \end{array}\right., 
\end{equation}
where
\begin{equation}
x=\frac{\lambda-d}{s}
\end{equation}
and
\begin{align}
a_0&=0.0406831, &
c_0&= 0.052035, \\
a_1&= 0.223201, &  
b_1&= -0.0551626, &
c_1&= 0.0062451, \\
d&=6000\,\AA, & s&=1000\,\AA.
\end{align}
The fit gives precision better than 10\%.

\end{document}